# Directional Scaling Symmetry of High-symmetry Two-dimensional Lattices

Longguang Liao and Zexian Cao

Institute of Physics, Chinese Academy of Sciences, Beijing 100190, China

E-mail: zxcao@iphy.ac.cn

Two-dimensional lattices provide the arena for many physics problems of essential importance, a non-trivial symmetry in such lattices will help to reveal the underlying physics. Whether there is a directional scaling for the 2D lattices is a longstanding puzzle. Here we report the discovery and proof of directional scaling symmetry for high symmetry 2D lattices, i.e., the square lattice, the equilateral triangular lattice and thus the honeycomb lattice, with the aid of the function $y = \arcsin(\sin(2\pi x n))$, where x is either the platinum number $\mu = 2 - \sqrt{3}$ or the silver number $\lambda = \sqrt{2} - 1$, which are related to the 12-fold and 8-fold quasiperiodic structures, respectively. The directions and the corresponding scaling factors for the symmetric scaling transformation are determined. The revealed scaling symmetry may have a bearing on the various physical problems modeled on 2D lattices, and the function adopted here can be used to generate quasiperiodic lattices with enumeration of lattice points. Our result is expected to initiate the search of directional scaling symmetry in more complicated geometries.

# I. Introduction

The square lattice and the equilateral triangular lattice, thus also the honeycomb lattice, are high-symmetry two-dimensional (2D) lattices. They play important roles in mathematics, physics, architectonics, arts and many other fields. For 2D lattices, the uniform scaling of the space, i.e., simultaneous dilation or contraction at two orthogonal directions with the same scale factors, will evidently preserve the character, or symmetry, of the patterns. A natural and perhaps also meaningful question may be raised: Is there any directional scaling symmetry for the 2D lattices of high symmetry that preserves the character of these lattices? Or in other words, is there any scaling transformation along a particular direction that brings a square (equilateral triangular) lattice into a square (equilateral triangular) lattice?

For the equilateral triangular lattice, if directional scaling is performed along any side of the unit triangle, i.e., along the <10>-directions, in the crystallographic nomenclature, then the contraction at any rate will never result in a triangular lattice. However, stretching along that direction with a scale factor $\gamma = 3$ results in a perfect equilateral triangular lattice, and the side length of the unit triangle in the resulting lattice is $\sqrt{3}$ times larger (Fig.1). A particular feature should be noticed that the neighborhood relation of the lattice points has been changed by this transformation. For example, in Fig.1a the lattice points (0, 1, 2, 3) form two unit triangles, $\Delta 013$ and $\Delta 023$, but in the resulting lattice in Fig.1b the two unit triangles formed by the corresponding lattice points are $\Delta 012$ and $\Delta 123$. The scale factor $\gamma = 3$ is the sole possibility of directional scaling symmetry for stretching along the side of the unit triangle in the case of equilateral triangular lattice. For scaling along the bisector line of a unit triangle, the solely possible directional scaling symmetry is the contraction with a scale factor $\gamma = 1/3$, which is in fact the inversed transformation of the one described above. This provides a trivial example of directional scaling symmetry. In the case of square lattice, directions along the side or the diagonal of the unit square don't exhibit any scaling symmetry.

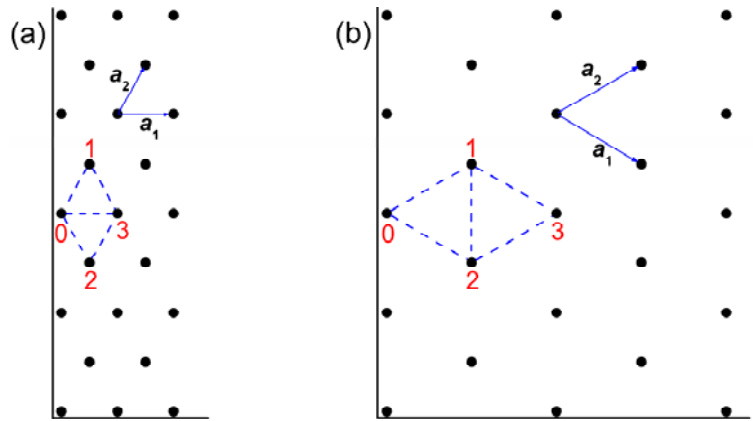

Fig.1. A trivial example of directional scaling symmetry for equilateral triangular lattice, which is achieved along any side of a unit triangle with a scale factor $\gamma = 3$. (a) The original lattice; (b) the transformation result of (a) along the connection line between points 0 and 3.

The lack of scaling symmetry along the most notable <10>-directions in the square and equilateral triangular lattices does not compulsively exclude the possibility of directional scaling symmetry along other directions. Rather, we may even wish that such a directional scaling symmetry, if there is any, can be achievable in principle with more scale factors. We see that if such a directional scaling symmetry can be proven to exist, and the corresponding transformation can be formulated, this will evidently promote our understanding of the structural properties of lattices, and provide helpful insight into problems involving lattices such as in statistical physics, condensed matter physics, quantum theory, and even number theory, etc.

In the effort of investigating the 1D incommensurate systems such as specified by the function $\cos(2\pi q n)$ [1-4], where is the integer and q is a Diophantine number such as the golden ratio, and 2D quasiperiodic structures [5-8], we came across to the question whether there is any directional scaling symmetry for the square lattice and the equilateral triangular lattice (hence also the honeycomb lattice). We found that the square lattice exhibits directional scaling symmetry along a direction at 22.5° with respect to the side of a unit square, with the drag center of scaling transformation falling on the lattice point, and the scale factor is $(3-2\sqrt{2})^k$, k=1,2,3…. In the case of equilateral triangular lattice, the directional scaling symmetry appears at the direction at 15° with respect to any side of the unit triangle, with the drag center of scaling transformation falling on the lattice point, and the scaling factor is $(7-4\sqrt{3})^k$, k=1, 2, 3…. A proof based on the function $y = \arcsin(\sin(2\pi xn))$, where x is the silver number [9-11] or the platinum number [11-13], which are respectively related to the 8-fold and 12-fold quasiperiodic structures, is presented.

**IIa. Directional scaling symmetry in equilateral triangular lattice**

The plot of the function $y = \sin n$, where the argument n is non-negative integer (the discussion below is also valid for negative integer, but it is not of concern here), is essentially different from that for $y = \sin x$, where x is real. This fact has been noticed and extensively studied by Strang [14,15] and Richert [16]. In studying the 1D incommensurate structures, we found that the function $y = \sin(2\pi\mu n)$, where $\mu = 2 - \sqrt{3}$ is the platinum number which is related to the dodecagonal quasiperiodic structure [8,11-13], reveals an interesting picture as illustrated in Fig.2a. In the boundary regions defined by $y = \pm 1$, the graph seems folding together, reminding us of Escher's paintings based on the concept of Poincaré disc. In the central region, however, the graph seems to display locally 12-fold rotational symmetry. This is quite reasonable since $\mu = 2-\sqrt{3}$ is the platinum number. If instead of $y = \sin(2\pi\mu n)$ we draw the function $y = \arcsin(\sin(2\pi\mu n))$, we see that the whole domain bounded by $y = \pm\pi/2$ is globally isometric, and the plot displays locally 12-fold rotational symmetry (in a not very strict sense), see Fig. 2b.

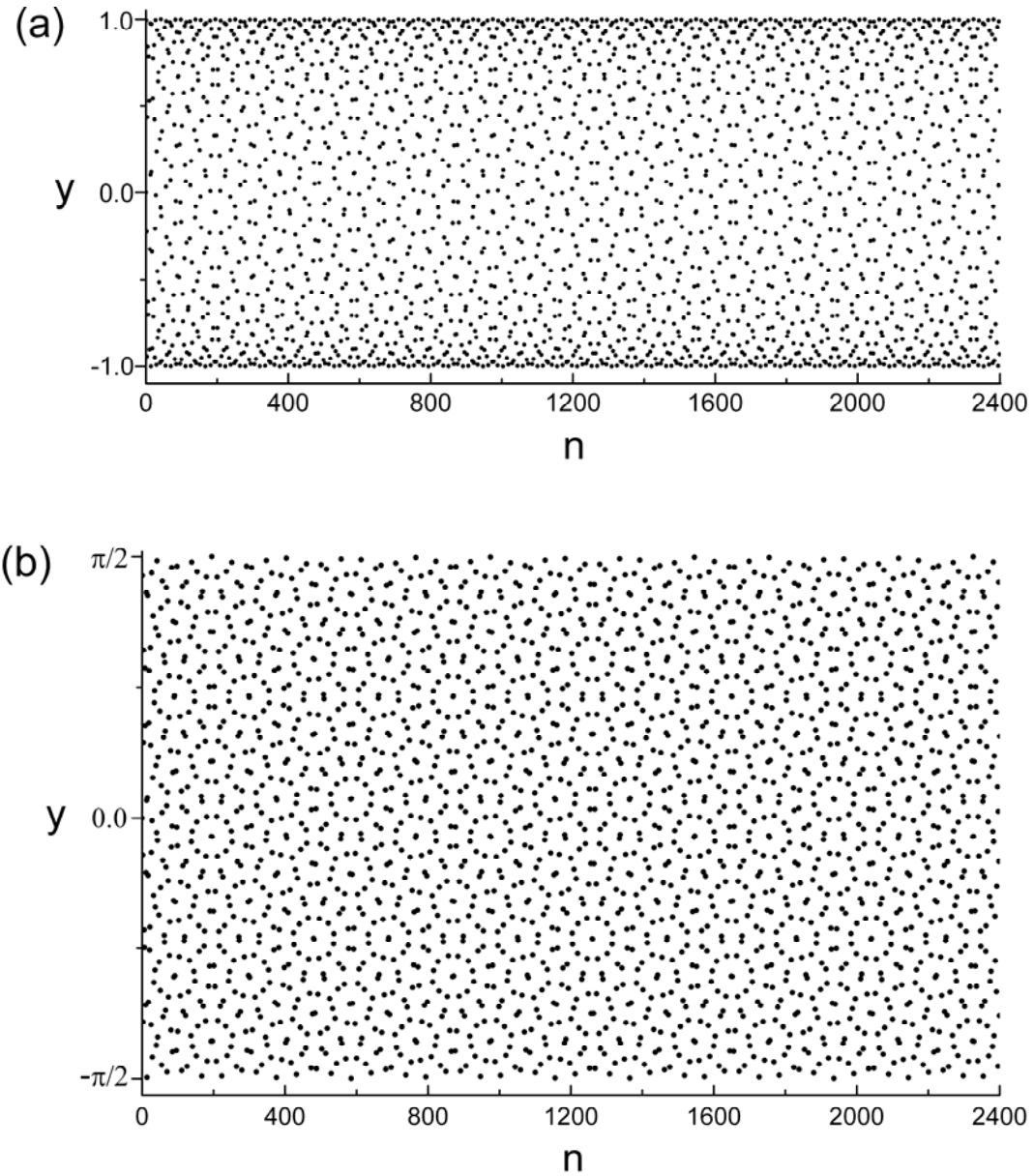

Fig.2. Plots of the sinusoidal function $y = \sin(2\pi\mu n)$ (a) and the arcsine function $y = \arcsin(\sin(2\pi\mu n))$ (b), where $\mu = 2 - \sqrt{3}$, and the argument n is non-negative integer.

Interestingly, the plot of the function $y = \arcsin(\sin(2\pi\mu n))$ in Fig.2b can be taken as a Moiré pattern, i.e., as superposition of two identical simpler lattices (see Fig.S1 in appendix). In fact, the function $y = \arcsin(\sin(2\pi\mu n))$ itself can be divided into two branches

$$\arcsin(\sin(2\pi\mu n)) = \begin{cases} 2\pi(n\mu - m), & \left(m - \tfrac{1}{4}\right) \le n\mu \le \left(m + \tfrac{1}{4}\right); \\ -2\pi(n\mu - m - \tfrac{1}{2}), & \left(m + \tfrac{1}{4}\right) < n\mu < \left(m + \tfrac{3}{4}\right). \end{cases} \qquad (1)$$

where both *m*, *n* are non-negative integer, and if $n\mu - [n\mu] \in [0, 3/4]$, then $m = [n\mu]$; if $n\mu - [n\mu] - 1 \in [-1/4, 0]$, then $m = [n\mu] + 1$. Here [x] denotes the truncation of the positive real number x. In the following the first branch in eq.(1) is referred to as the ascending branch, as points generated by this branch fall on the ascending part of the graph for $y = \arcsin(\sin(x))$ (see Fig.S2 in appendix), and the second branch is accordingly referred to as the descending branch. The plot of only the ascending branch results in Fig.3a (for comparison of the two branches, see Fig.S1 in appendix). From Fig. 3a we can readily find that the plot of the ascending branch constitutes an oblique 2D lattice. So does the plot of the descending branch. In fact, with a proper ratio of the longitudinal scale to the transverse scale the unit triangle in Fig.3a can be made to have roughly three equal sides, thus the lattice is approximately an equilateral triangular lattice (to be further discussed below).

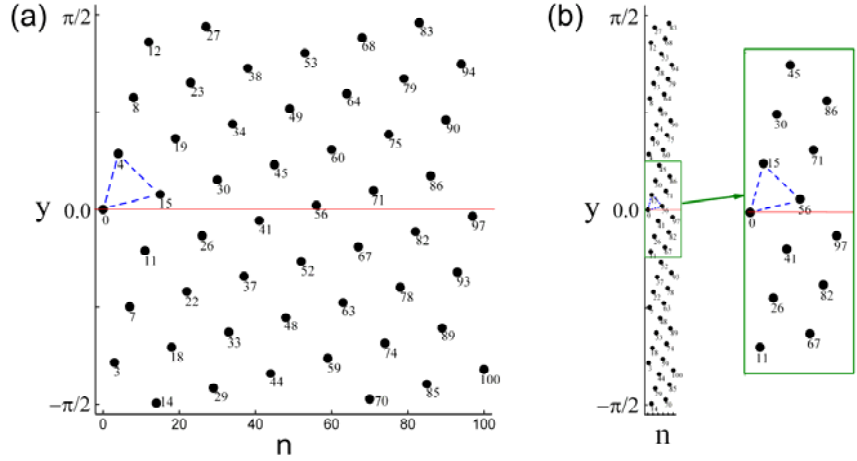

Fig.3. (a) Plot of the ascending branch of the function $y = \arcsin(\sin(2\pi\mu n))$, where $\mu = 2 - \sqrt{3}$, and n is non-negative integer; (b) The result of scaling along the horizontal axis with a scale factor of $\sim 7 - 4\sqrt{3}$. Points are indexed with the corresponding argument n.

If we compress Fig.3a along the horizontal axis in a continuous way, the approximate equilateral triangle lattice will at first be distorted, and then, when the scale factor comes to a proper value ($\sim 7 - 4\sqrt{3}$), the shape of the lattice will again recover, as illustrated in Fig.3b. This scenario can be repeated infinitely. More importantly, after each contraction, the unit triangle in the lattice can be brought closer to a rigorously equilateral triangle, in the sense that the side lengths suffer from a less relative deviation. And it can be proven that in the extreme case when the ratio of longitudinal scale to transverse scale approaches vanishingly small, the unit triangle turns into a rigorously equilateral triangle (see detailed proof in appendix). Notice that the transformation changes the neighborhood relation that, for instance, in Fig.3a the two unit triangles anchored to the point n=0 are $\Delta 0-4-15$ and $\Delta 0-11-15$, whereas after the transformation, the two unit triangles anchored to the point n=0 are $\Delta 0-15-56$, and $\Delta 0-41-56$ (Fig.3b).

Thus this manipulation leads us to the discovery that there exists directional scaling symmetry for an equilateral triangular lattice, which is a scaling transformation, setting the drag point on an arbitrary lattice point, along the direction at 15° with respect to the side of the unit triangle, and the scale factor is $7-4\sqrt{3}$. The ratio of side lengths involved in this transformation is $2-\sqrt{3}$. Such a scaling transformation can be performed repeatedly. This directional scaling symmetry for equilateral triangular lattice specified above can be easily checked (see detailed proof in appendix).

By the way, the equilateral triangular lattice is the superposition of a honeycomb lattice and a $\sqrt{3}$ times larger equilateral triangular lattice. Taking the lattice in Fig.3a as an equilateral triangular lattice, the index in the plot helps to specify the points to be removed so as to obtain a honeycomb lattice from the parent triangular lattice (The rules of doing this are clarified in appendix). Obviously, the honeycomb lattice has also directional scaling symmetry, and the scale factor and the ratio of side lengths for hexagons before and after the transformation are $7-4\sqrt{3}$ and $2-\sqrt{3}$, respectively.

The drag point is set on the center of an arbitrary unit hexagon, and the direction is at 15° with respect to the side of the hexagon. More interestingly, when a honeycomb is obtained after scaling along that particular direction, the center of the unit hexagon remains the center of the unit hexagon in the resulting lattice. The honeycomb lattice and equilateral triangular lattice share the same directional scaling symmetry may arise from the fact that honeycomb lattice is dual ( reciprocal) to the equilateral triangular lattice.

## IIb. Directional scaling symmetry in square lattice

With the silver number $\lambda = \sqrt{2} - 1$ instead of $\mu = 2 - \sqrt{3}$, which is related to the octagonal quasiperiodical structure [7, 9-11], we obtain an interesting plot of the function $y = \sin(2\pi\lambda n)$ (Fig.4a) in analog to Fig.2a. Going one step further, we draw the plot of the function $y = \arcsin(\sin(2\pi\lambda n))$, which is globally isometric, and displays locally 8-fold rotational symmetry (in a not very strict sense), see Fig. 4b.

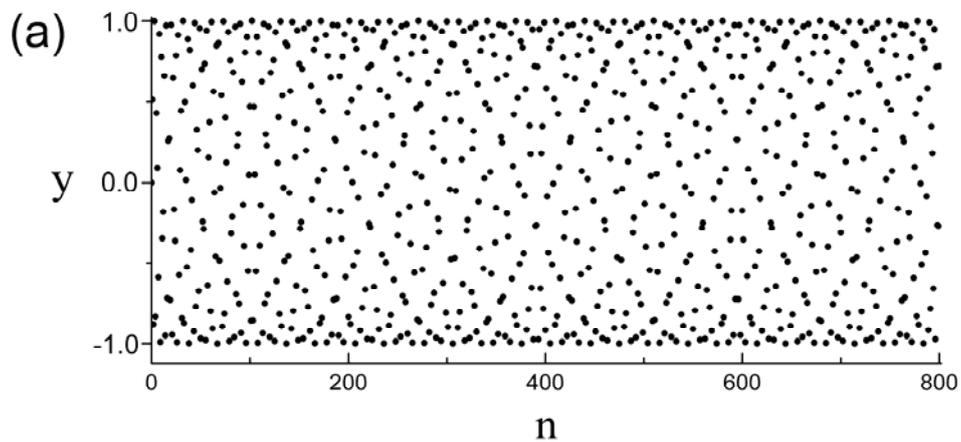

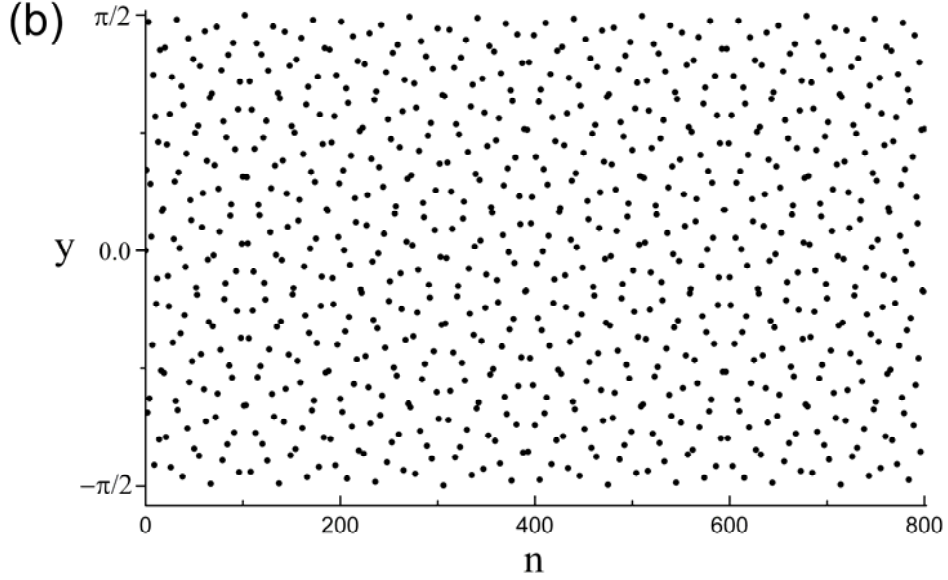

Fig.4. Plots of the sinusoidal function $y = \sin(2\pi\lambda n)$ (a) and the arcsine function $y = \arcsin(\sin(2\pi\lambda n))$ (b), where $\lambda = \sqrt{2} - 1$, and the argument n is non-negative integer.

Again, the plot in Fig.4b can be taken as a Moiré pattern formed by the superposition of two identical simpler lattices (see Fig.S3 in appendix). Accordingly, the function $y = \arcsin(\sin(2\pi\lambda n))$ can be separated into two branches

$$\arcsin(\sin(2\pi\lambda n)) = \begin{cases} 2\pi(n\lambda - m), & \left(m - \tfrac{1}{4}\right) \leq n\lambda \leq \left(m + \tfrac{1}{4}\right); \\ -2\pi(n\lambda - m - \tfrac{1}{2}), & \left(m + \tfrac{1}{4}\right) < n\lambda < \left(m + \tfrac{3}{4}\right). \end{cases} \quad (2)$$

Where *m*, *n* are non-negative integers, and if $n\lambda - [n\lambda] \in [0, 3/4]$, then $m = [n\lambda]$; if $n\lambda - [n\lambda] - 1 \in [-1/4, 0]$, then $m = [n\lambda] + 1$. As above, the first branch is referred to as the ascending branch of the function, and the second branch is referred as the descending branch. Thus the plot of $y = \arcsin(\sin(2\pi\lambda n))$ can be taken as the Moiré pattern formed by the overlapping plots for its ascending branch and descending branch (see Fig.S3 in appendix).

The ascending branch of the function $y = \arcsin(\sin(2\pi\lambda n))$ is plotted in Fig.5a (For comparison of the two branches, see Fig.S3 in appendix). One can easily check that the points in Fig.5a form a square lattice, in an approximate sense, when a proper ratio of longitudinal scale to transverse scale is chosen (see detailed proof in appendix).

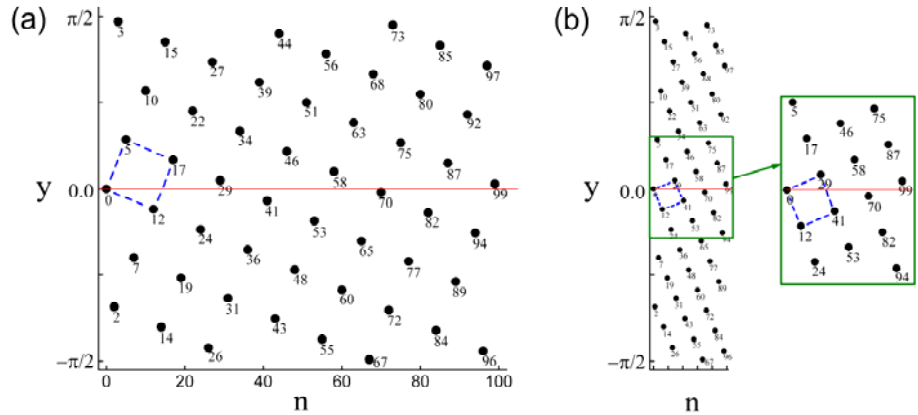

Fig.5. (a) Plot of the ascending branch of the function $y = \arcsin(\sin(2\pi\lambda n))$, where $\lambda = \sqrt{2} - 1$, and n is non-negative integer; (b) The result of scaling along the horizontal axis with scale factor $\sim 3 - 2\sqrt{2}$. Points are indexed with the corresponding argument n.

If Fig.5a is compressed along the horizontal axis, the shape of approximate square unit will at first be distorted, then, when the scale factor comes to a value $\sim 3 - 2\sqrt{2}$, the shape of the lattice will be recovered, as illustrated in Fig.5b. This operation can be performed repeatedly. After each contraction, the approximate unit square gets closer to a rigorous square (see detailed proof in appendix). It can be proven that the approximate unit square turns into a rigorous square when the ratio of longitudinal scale to transverse scale becomes vanishing small (see detailed proof in appendix). Notice that the neighborhood relation of points in the lattice has been changed by contraction. For example, the unit square, anchored to the original point 0, is (□0-5-12-17) in Fig.5a, but after the contraction it is the square □0-12-29-41, see Fig.5b. Moreover, the unit square is also rotated by 45° by the transformation.

Thus this manipulation leads us to the discovery that directional scaling symmetry exists for the square lattice, which is along the direction at 22.5° with respect to any side of a unit square, and the scaling factor is $3 - 2\sqrt{2}$. The ratio of the side lengths of the unit squares before and after transformation is $\sqrt{2} - 1$ (see detailed proof in appendix).

**III. Summary**

By using the arcsine functions $y = \arcsin(\sin(2\pi xn))$, where the parameter x is either the platinum number $\mu = 2 - \sqrt{3}$ or the silver number $\lambda = \sqrt{2} - 1$, we found and proved the existence of directional scaling symmetry for the equilateral triangular lattice (thus also the honeycomb lattice), and the square lattice. With the drag center set on a lattice point, in the case of equilateral triangular lattice, the direction of scaling symmetry is at 15° with regard to the side of the unit triangle, and the scale factor is $7 - 4\sqrt{3}$, while in the case of square lattice, the direction of scaling symmetry is at 22.5° with regard to the side of the unit square triangle, and the scale factor is $3 - 2\sqrt{2}$. In both cases the scaling transformation can be performed repeatedly. The proof of the directional scaling symmetry for 2D lattices may suggest directional scaling symmetry for 3D cubic and rhombic lattices, which is an open question that we will bet on a positive answer. With the current work we want to call the attention to this directional scaling symmetry for the equilateral triangular lattice and square lattice, which is expected to have some impact on the various physics problems, particularly in statistical physics, condensed matter physics, quantum field theory, etc., modeled on such lattices. Moreover, the method of proof involving the trigonometric functions with Diophantine numbers in the argument, and approaching a property of the rigorous symmetry lattice from the approximate lattices, is new and inspiring. In fact, such a function can be used to generate quasiperiodic lattices with enumeration of the lattice points, which is very helpful for the calculation, of energy bands, say, for quasicrystals. These are a few exciting topics of future research for applied mathematics and physics.

**Appendix**

**Contents**

I. Figures S1-S3

   Fig.S1. Plot of both branches of the function $y = \arcsin(\sin(2\pi\mu n))$, $\mu = 2 - \sqrt{3}$;

   Fig.S2: Plot of the function $y = \arcsin(\sin(x))$, x is real;

   Fig.S3: Plot of both branches of the function $y = \arcsin(\sin(2\pi\lambda n))$, $\lambda = \sqrt{2} - 1$.

II. Platinum number, dodecanacci sequence, and proof of directional scaling symmetry for equilateral triangular lattice

III. Silver number, octonacci sequence, and proof of directional scaling symmetry for square lattice

IV. Obtaining honeycomb lattice from the equilateral triangular lattice

References

# I. Figures S1-S3

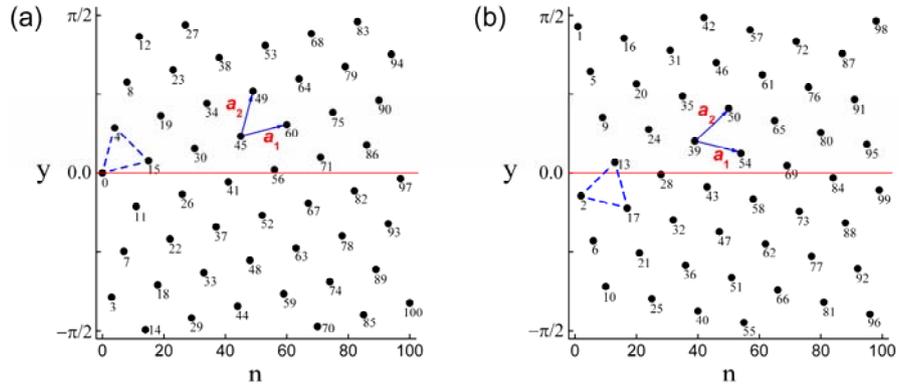

Fig.S1. Plot of the two branches of the function $y = \arcsin(\sin(2\pi\mu n))$, $\mu = 2-\sqrt{3}$, both illustrating an approximate equilateral triangular lattice. (a) The ascending branch; (b) the descending branch. Index of points in the figure is the corresponding value of the argument n.

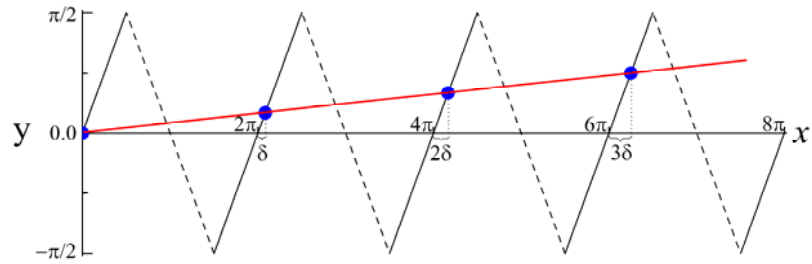

Fig.S2. Plot of the function $y = \arcsin(\sin(x))$, where x is real, divided into the ascending branch (solid line) and the descending branch (dashed line). Joint points of a line with the individual branch, ascending or descending, are separated at equal distance.

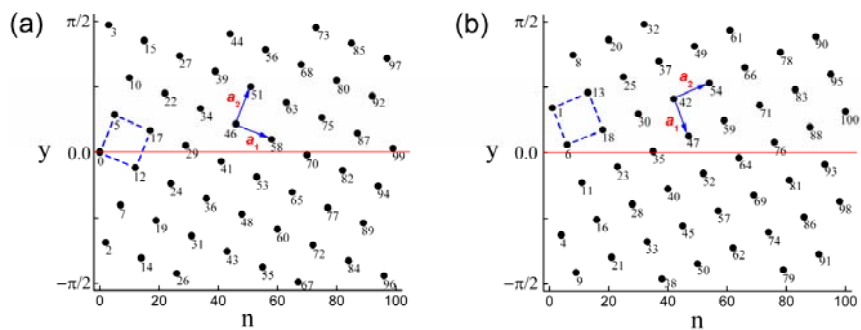

Fig.S3. Plot of the two branches of the function $y = \arcsin(\sin(2\pi\lambda n))$, $\lambda = \sqrt{2} - 1$, both illustrating an approximate square lattice. (a) The ascending branch; (b) the descending branch. Index of points in the figure is the corresponding value of the argument n.

## II. Platinum number, dodecanacci sequence, and proof of directional scaling symmetry for equilateral triangular lattice

For the proof of directional scaling symmetry for equilateral triangular lattice, some preparatory work has to be done on the basis of platinum number and dodecanacci sequence.

The quadratic equation $x^2 - 4x + 1 = 0$ has two real roots, one is the platinum number [A1, A2] $\rho = 2 + \sqrt{3}$, the other is its reciprocal $\mu = (2 - \sqrt{3}) = \tan(\pi/12)$. Both roots are characteristic numbers for the dodecagonal quasiperiodic structure [A3]. Table 1 lists the k-th power of $\mu$, which will be used in the proof.

**Table 1.** The $k$-th power of $\mu = (2 - \sqrt{3})$

| | $\mu^k$ |
|---|---|
| 0 | $\mu^0 = 1 - 0\sqrt{3} = 1$ |
| 1 | $\mu^1 = 2 - 1\sqrt{3} = 1\mu - 0$ |
| 2 | $\mu^2 = 7 - 4\sqrt{3} = 4\mu - 1$ |
| 3 | $\mu^3 = 26 - 15\sqrt{3} = 15\mu - 4$ |
| 4 | $\mu^4 = 97 - 56\sqrt{3} = 56\mu - 15$ |
| 5 | $\mu^5 = 362 - 209\sqrt{3} = 209\mu - 56$ |
| 6 | $\mu^6 = 1351 - 780\sqrt{3} = 780\mu - 209$ |
| … | …… |
| k | $\mu^k = I_k - D_k\sqrt{3} = D_k\mu - D_{k-1}$ |

From the $\mu$-representation of the powers of $\mu$, we obtain two number sequences: one is 0, 1, 4, 15, 56, 209, 780,…; the other is 1, 2, 7, 26, 97, 362, 1351,…. These two sequences are dubbed the dodecanacci sequence by some researchers [A3].

In order to facilitate the discussion below, we denote the first sequence as $D$, the second sequence as $I$, and the $k$-th items of these two sequences as $D_k$ and $I_k$, respectively. The two dodecanacci sequences $D$ and $I$ have the following simple iterative relations

$$D_{k+1} = 4D_k - D_{k-1}, \quad D_0 = 0, \ D_1 = 1; \qquad (1)$$

$$I_{k+1} = 4I_k - I_{k-1}, \quad I_0 = 1, \ I_1 = 2; \qquad (2)$$

This is to say that they are generated from the same iteration process with different initial items. For both sequences $D$ and $I$, the ratio of an item over the preceding one approaches the platinum number $\rho = 2 + \sqrt{3}$ as $k \to \infty$.

From the expression in table 1,

$$\mu^k = D_k \mu - D_{k-1} \qquad (3)$$

it follows that $D_k \mu - D_{k-1} > 0$, and it vanishes as $k \to \infty$. The number sequence $D$ also satisfies the following equality

$$D_{k+1}^2 - 4D_k D_{k+1} + D_k^2 = 1 \qquad (4)$$

The number $\mu = 2-\sqrt{3}$, which is irrational, can be expressed as a periodic, infinitely continued fraction [A4], i.e., $\mu = 2-\sqrt{3} = [0, 3, 1, 2, 1, 2, 1, 2, 1, 2, ...]$. If the continued fraction is cut off at the $l$-th order, then we can get a rational approximation $\tilde{\mu}_l = A_l / B_l$ [A4-A6] to this irrational number. For the case of $l = 2k$, $k \geq 0$, the 2k-th order approximation can be expressed as

$$\tilde{\mu}_{2k} \approx \frac{D_k}{D_{k+1}}. \qquad (5)$$

Obviously, as $k$ increases, the rational approximation approaches $\mu = 2-\sqrt{3}$. Similarly,

for $l = 2k+1, k \geq 0$, the (2k+1)-th order rational approximation can be expressed as

$$\tilde{\mu}_{2k+1} \approx \frac{D_k}{D_{k+1}}.$$

Now we turn to the existence proof of directional scaling symmetry for the equilateral triangular lattice along the following route:

1) Showing that the points generated by a branch of the function $y = \arcsin(\sin(2\pi\mu n))$ form an oblique lattice;

2) Showing that in the plot of the ascending branch, the unit triangle $D_0 D_k D_{k+1}$, which is possible at a proper ratio of the longitudinal scale to the transverse scale, deviates less and less from a rigorous equilateral triangle with increasing k. Thus at $k \to \infty$, the lattice turns into an equilateral triangular lattice, and the scaling process along the horizontal axis in the extreme reveals the directional scaling symmetry for the equilateral triangular lattice;

3) Working out the parameters of directional scaling symmetry.

Unless in the cases that y assumes a value of $\pi/2$, the points generated by the ascending (descending) branch (defined in eq.(1) in the text) of the function $y = \arcsin(\sin(2\pi\mu n))$, $\mu = 2 - \sqrt{3}$, fall on the ascending (descending) branch of the function $y=\arcsin(\sin x)$, and those points on a line are separated at equal distance. Based on these facts, the points generated by a given branch of the function $y = \arcsin(\sin(2\pi\mu n))$ form an oblique lattice (Fig.S1). This can be confirmed by calculating the distances and orientations of neighboring points with regard to a given point. The calculation also indicates that the lattices in Fig.S1 are approximately an equilateral triangular lattice (see discussion below).

**Theorem 1:** Points referred to the number in the series $D_k, k \geq 2$, appear in the ascending branch of the function $y = \arcsin(\sin(2\pi\mu n))$, $\mu = 2 - \sqrt{3}$, and satisfy the condition that if $0 < n < D_k$, then $|y(n)| > |y(D_k)|$. Points indexed by the number series $I_k, k \geq 2$, also appear in the ascending branch, and they approach the horizontal axis with increasing k.

**Proof:** The rational approximation to $\mu = 2-\sqrt{3}$ obtained with the method of continued fraction, $\tilde{\mu}_{2k} \approx \dfrac{D_k}{D_{k+1}}$, is a best approximation to this irrational number [A4,A6]. The best approximation of a real number $x$ is the rational fraction $A/B$ such that it has $|B'x - A'| > |Bx - A|$ for $0 < B' < B$ and $A'/B' \neq A/B$ [A4,A6]. Here A, B, A', B' are all positive integers.

According to eq.(1) in the text, the point labeled with 0 is one belonging to the ascending branch of the function $y = \arcsin(\sin(2\pi\mu n))$, while points 1 and 2 are not. Point 3 also belongs to the ascending branch since $|3\mu - 1| \leq 1/4$, so does point 4 since $|4\mu - 1| \leq 1/4$. Based on the theory of best approximation for a real number, it has $|D_k \mu - D_{k-1}| < |D_2 \mu - D_1| < 1/4$ for $k \geq 3$. Consequently, the points labeled by $D_k$ for $k \geq 2$ appear in the ascending branch. This means that the point specified by $D_k$ approaches the horizontal axis with increasing k.

Hence, for points labeled by a number $n$ in the ascending branch of the function $y = \arcsin(\sin(2\pi\mu n))$, the following relation always holds: $2\pi|n\mu - m| > 2\pi|D_k \mu - D_{k-1}|$, or $|y(n)| > |y(D_k)|$, for any $0 < n < D_k$, n and m are any positive integers.

The lattice defined by the points in one branch of the function $y = \arcsin(\sin(2\pi\mu n))$ can be only an approximate equilateral triangular lattice. To see this point, one needs check the side lengths of the triangle defined by points $D_0$ (0) and other two points $D_k$ and $D_{k+1}$, for instance $D_2$ (4) and $D_3$ (15), which forms a unit triangle in the lattice (Fig.S1). Let's compare the distances of the points $D_k$ and $D_{k+1}$ to the origin $D_0$ (0), under the condition that the ratio of the longitudinal scale to the transverse scale is set at $\gamma = \gamma_k$ such that the distance of $D_k$ to the origin is equal to that between $D_k$ and $D_{k+1}$, i.e., $d_k(D_0 D_k) = d_k(D_k D_{k+1})$. Since

$$d_k^2(D_0 D_k) = (D_k \gamma_k)^2 + \left[2\pi(D_k \mu - D_{k-1})\right]^2 = D_k^2 \gamma_k^2 + 4\pi^2 \mu^{2k} \qquad (6)$$

$$d_k^2(D_k D_{k+1}) = (D_{k+1} - D_k)^2 \gamma_k^2 + \left[2\pi(D_{k+1}\mu - D_k) - 2\pi(D_k \mu - D_{k-1})\right]^2$$
$$= (D_{k+1} - D_k)^2 \gamma_k^2 + 4\pi^2 \left(\mu^{2k+2} - 2\mu^{2k+1} + \mu^{2k}\right) \qquad (7)$$

and

$$d_k^2(D_0 D_{k+1}) = (D_{k+1}\gamma_k)^2 + \left[2\pi(D_{k+1}\mu - D_k)\right]^2 = D_{k+1}^2 \gamma_k^2 + 4\pi^2 \mu^{2k+2} \qquad (8)$$

From eqs.(6) and (7), and using the iterative relations (1) and (3), and with the fact $\mu = 2 - \sqrt{3}$, we get

$$\gamma_k^2 = \frac{4\sqrt{3}\pi^2 \mu^{2k+1}}{\left(8D_k^2 - 6D_{k-1}D_k + D_{k-1}^2\right)} \qquad (9)$$

Substituting this $\gamma_k^2$ into eqs. (6) and (8), we get

$$d_k^2(D_0 D_k) = d_k^2(D_k D_{k+1}) = 4\pi^2 \mu^{2k} \frac{\left[(5 + 2\sqrt{3})D_k^2 - 6D_{k-1}D_k + D_{k-1}^2\right]}{\left(8D_k^2 - 6D_{k-1}D_k + D_{k-1}^2\right)} \qquad (10)$$

$$d_k^2(D_0 D_{k+1}) = 8\pi^2 \mu^{2k+1} \frac{\left[(\sqrt{3}+2)D_k - D_{k-1}\right]}{(2D_k - D_{k-1})} = 8\pi^2 \mu^{2k} \frac{[D_k - \mu D_{k-1}]}{(2D_k - D_{k-1})} \qquad (11)$$

The difference between $d_k^2(D_0 D_k)$ and $d_k^2(D_0 D_{k+1})$ is thus given by

$$\delta_k = d_k^2(D_0 D_k) - d_k^2(D_0 D_{k+1}) = \gamma_k^2 \qquad (12)$$

This is to say that the unit triangle in the lattice defined by the ascending branch of the function $y = \arcsin(\sin(2\pi\mu n))$ is not a rigorous equilateral triangle. It can, however, be a rigorous equilateral triangle in the extreme case when $\gamma_k$ approaches zero. The same conclusion can be drawn for the descending branch.

Now we further investigate the case that the triangle ($D_0$, $D_{k+1}$, $D_{k+2}$), for instance (0,15, 56) in Fig.3b, is a unit triangle in the lattice. Again, let's check the distance between point $D_{k+1}$ and point $D_{k+2}$, and their distances to the origin.

Let the ratio of the longitudinal scale to the transverse scale now be $\gamma = \gamma_{k+1}$ when the distance from $D_{k+1}$ to the origin is set equal to the distance between $D_{k+1}$ and $D_{k+2}$, i.e., $d_{k+1}(D_0 D_{k+1}) = d_{k+1}(D_{k+1} D_{k+2})$. Following the discussion concerning $\gamma_k^2$ in the above paragraph, we can get

$$\gamma_{k+1}^2 = \frac{4\sqrt{3}\pi^2 \mu^{2k+3}}{\left(105 D_k^2 - 58 D_{k-1} D_k + 8 D_{k-1}^2\right)} \tag{13}$$

and

$$\begin{aligned} d_{k+1}^2(D_0 D_{k+1}) &= d_{k+1}^2(D_{k+1} D_{k+2}) \\ &= 4\pi^2 \mu^{2k+2} \frac{\left[\left(57 + 32\sqrt{3}\right) D_k^2 + \left(-34 - 16\sqrt{3}\right) D_{k-1} D_k + \left(5 + 2\sqrt{3}\right) D_{k-1}^2\right]}{\left(105 D_k^2 - 58 D_{k-1} D_k + 8 D_{k-1}^2\right)} \end{aligned} \tag{14}$$

$$d_{k+1}^2(D_0 D_{k+2}) = 8\pi^2 \mu^{2k+2} \frac{\left[\left(\sqrt{3}+2\right) D_k - D_{k-1}\right]}{(7 D_k - 2 D_{k-1})} = 8\pi^2 \mu^{2k+1} \frac{[D_k - \mu D_{k-1}]}{(7 D_k - 2 D_{k-1})} \tag{15}$$

Thus finally we get,

$$\delta_{k+1} = d_{k+1}^2(D_0 D_{k+1}) - d_{k+1}^2(D_0 D_{k+2}) = \gamma_{k+1}^2. \tag{16}$$

This is to say that an approximate triangular lattice appears once again when the ratio of the longitudinal scale to the transverse scale is set at $\gamma = \gamma_{k+1}$, and $\delta_{k+1} = \gamma_{k+1}^2$.

The deviation of the unit triangle ($D_0$, $D_k$, $D_{k+1}$), at $\gamma = \gamma_k$, from a rigorous equilateral triangle can be evaluated by the parameter $\Delta_k = \dfrac{d_k^2(D_0 D_k) - d_k^2(D_0 D_{k+1})}{d_k^2(D_0 D_k)}$.

From eqs. (10) and (12) we get

$$\Delta_k = \frac{\left[\sqrt{3}\left(2 - \sqrt{3}\right)\right]}{\left[\left(5 + 2\sqrt{3}\right) D_k^2 - 6 D_{k-1} D_k + D_{k-1}^2\right]} \tag{17}$$

And from eqs.(14) and (16), it has

$$\Delta_{k+1} = \frac{\sqrt{3}\left(2 - \sqrt{3}\right)}{\left(57 + 32\sqrt{3}\right) D_k^2 + \left(-34 - 16\sqrt{3}\right) D_{k-1} D_k + \left(5 + 2\sqrt{3}\right) D_{k-1}^2} \tag{18}$$

Since $D_k/D_{k-1} \geq \left(2 + \sqrt{3}\right)$, it can be easily shown that $\Delta_k / \Delta_{k+1} < 1$. This is to say that the deviation from a rigorous equilateral triangle at $\gamma = \gamma_{k+1}$ is less than that at $\gamma = \gamma_k$. At $k \to \infty$, $\Delta_k \to 0$, and the points obtained from the ascending branch of the function $y = \arcsin(\sin(2\pi\mu n))$ form a rigorous equilateral triangular lattice. The same conclusion can be drawn for the descending branch.

The scaling at $k \to \infty$ is then operated on an equilateral triangular lattice, which reveals the presence of the directional scaling symmetry (along the horizontal direction in Fig.S1) for the equilateral triangular lattice.

The connection line between point 0 and point 15 in Fig.3a, and the connection line between point 0 and point 56 in Fig.3b, are roughly at 15° ($\pi/12$) respect to the horizontal axis. The calculation below points to the conclusion that the direction of scaling symmetry for the equilateral triangular lattice, going through a lattice point, lies at 15° with regard to any side of the unit triangle.

Combining with eq.(8), the equality $y(D_{k+1}) = 2\pi(D_{k+1}\mu - D_k) = 2\pi\mu^{k+1}$, and the fact that $\sin(15°) = \frac{\sqrt{6}-\sqrt{2}}{4}$, we obtain, for $\gamma = \gamma_k$,

$$y^2(D_{k+1}) - d_k^2(D_0 D_{k+1})\sin^2(15°) = 2\pi^2\mu^{2k+2}\left[\frac{(2-\sqrt{3})D_k - D_{k-1}}{2D_k - D_{k-1}}\right] \geq 0 \quad (19)$$

At $k \to \infty$, the equality holds. Thus the direction of scaling symmetry for the equilateral triangular lattice is, setting the drag point at a lattice point, at 15° with respect to the side of the unit triangle.

Combining eq.(9) and eq.(13), we can obtain the ratio of $\gamma_{k+1}^2$ over $\gamma_k^2$,

$$\frac{\gamma_{k+1}^2}{\gamma_k^2} = \frac{\mu^2\left(8D_k^2 - 6D_{k-1}D_k + D_{k-1}^2\right)}{\left(105D_k^2 - 58D_{k-1}D_k + 8D_{k-1}^2\right)} \quad (20)$$

At $k \to \infty$, $D_k/D_{k-1} \to 2+\sqrt{3}$, we get

$$\lim_{k \to \infty} \frac{\gamma_{k+1}}{\gamma_k} = \mu^2 = 7 - 4\sqrt{3} \quad (21)$$

Which is the scale factor involved in the directional scaling symmetry for the equilateral triangular lattice. Accordingly, the ratio of the side lengths of the unit triangles before and after the scaling is $2-\sqrt{3}$. We notice that the scale factor is the squared ratio of side lengths of unit triangles involved in the directional scaling symmetry for equilateral triangular lattice. This is a notable feature of directional scaling symmetry.

In summary, for the equilateral triangular lattice, scaling symmetry appears at the direction, going through a lattice point, at 15° with respect to the side of the unit triangle, the scale factor is $7-4\sqrt{3}$, and the side length of the unit triangle is scaled by $\mu = 2-\sqrt{3}$. Such a deflation can be performed repeatedly.

### III. Silver number, octonacci sequence, and proof of directional scaling symmetry for square lattice

The proof of directional scaling symmetry for the square lattice follows the same procedure as for equilateral triangular lattice.

The quadratic equation $x^2 - 2x - 1 = 0$ has two roots, one is the silver number [A3, A7, A8] $\sigma = \sqrt{2} + 1$, the other is $-\sqrt{2} + 1$. Noticing that $\lambda = \sigma^{-1} = (\sqrt{2} - 1) = \tan(\pi/8)$, thus we see that both roots are characteristic numbers for the octagonal quasiperiodic structure [A3]. Table 2 lists the k-th power of $\lambda$.

**Table 2.** The k-th power of $\lambda$

| | $\lambda^k$ |
|---|---|
| 0 | $\lambda^0 = (-1)^{0+1}(0\sqrt{2} - 1)$ |
| 1 | $\lambda^1 = (-1)^{1+1}(1\sqrt{2} - 1) = (-1)^{1+1}(1\lambda - 0)$ |
| 2 | $\lambda^2 = (-1)^{2+1}(2\sqrt{2} - 3) = (-1)^{2+1}(2\lambda - 1)$ |
| 3 | $\lambda^3 = (-1)^{3+1}(5\sqrt{2} - 7) = (-1)^{3+1}(5\lambda - 2)$ |
| 4 | $\lambda^4 = (-1)^{4+1}(12\sqrt{2} - 17) = (-1)^{4+1}(12\lambda - 5)$ |
| 5 | $\lambda^5 = (-1)^{5+1}(29\sqrt{2} - 41) = (-1)^{5+1}(29\lambda - 12)$ |
| 6 | $\lambda^6 = (-1)^{6+1}(70\sqrt{2} - 99) = (-1)^{6+1}(70\lambda - 29)$ |
| ... | ...... |
| k | $\lambda^k = (-1)^{k+1}\left(O_k\sqrt{2} - J_k\right) = (-1)^{k+1}(O_k\lambda - O_{k-1})$ |

From the powers of $\lambda$, we obtain two number sequences: one is 0, 1, 2, 5, 12, 29, 70,…; the other is 1, 1, 3, 7, 17, 41, 99,…. These two sequences are dubbed the octonacci sequences [A3,A7]. In order to facilitate the following discussion, we denote the first sequence with $O$, the second with $J$, and the $k$-th items of these two sequences as $O_k$ and $J_k$, respectively. The two octonacci sequences $O$ and $J$ have the same simple iterative relation

$$O_{k+1} = 2O_k + O_{k-1}, \quad O_0 = 0, \ O_1 = 1; \tag{22}$$

$$J_{k+1} = 2J_k + J_{k-1}, \quad J_0 = 1, \ J_1 = 1; \tag{23}$$

This is to say that they can be generated from the same iteration process with different initial items. For both sequences $O$ and $J$, the ratio of an item over the preceding one approaches the silver number $\sigma = \sqrt{2}+1$ at $k \to \infty$. The number sequences O and J have the relation

$$J_k = O_k + O_{k-1}, \text{ or } J_k = O_{k+1} - O_k, \quad (k \geq 1). \tag{24}$$

Then, the k-th power of $\lambda$ can be expressed as

$$\lambda^k = (-1)^{k-1}\left(O_k\sqrt{2} - J_k\right) = (-1)^{k-1}\left(O_k\lambda - O_{k-1}\right) \tag{25}$$

Combining eqs.(24) and (25), we get

$$J_k\lambda - J_{k-1} = (-1)^k \lambda^{k-1}(1-\lambda) \tag{26}$$

From eq.(25), we see that the absolute value $|O_k\lambda - O_{k-1}|$ vanishes at $k \to \infty$.

The octonacci sequence $O$ satisfies the following relation,

$$\left[O_{k+1}^2 - O_k O_{k+2}\right] = \left[O_{k+1}^2 - 2O_k O_{k+1} - O_k^2\right] = (-1)^k \tag{27}$$

The number $\lambda$, which is irrational, can be expressed as a periodic, infinitely continued fraction, $\lambda = \sqrt{2}-1 = [0, 2, 2, 2, 2, 2, 2, ...]$. If the continued fraction is cut off at the *k-th* order, then we can get a rational approximation $\tilde{\lambda}_k = O_k / O_{k+1}$ [A4, A6] for $\lambda = \sqrt{2}-1$, the first item is defined as the 0-th order. With increasing k, the approximation approaches $\lambda = \sqrt{2}-1$.

Just as in the case of function $y = \arcsin(\sin(2\pi\mu n))$, $\mu = 2-\sqrt{3}$, unless that y assumes a value of $\pi/2$, the points generated by the ascending (descending) branch (defined in eq.(2) in the text) of the function $y = \arcsin(\sin(2\pi\lambda n))$, $\lambda=\sqrt{2}-1$, fall on the ascending (descending) branch of the function $y=\arcsin(\sin x)$, and the points on a line are separated at equal distance. Based on these facts, the points generated by a branch of the function $y = \arcsin(\sin(2\pi\lambda n))$ form an oblique lattice (Fig.S3). This can be further confirmed by calculating the distances and orientations of neighboring points with regard to a given point. The calculation also indicates that the lattices in Fig.S3 are approximately a square lattice (see discussion below).

**Theorem 2:** Points referred to the number in the series $O_k$, $k \geq 2$, appear in the ascending branch of the function $y = \arcsin(\sin(2\pi\lambda n))$, $\lambda=\sqrt{2}-1$, and satisfy the condition that if $0 < n < O_k$, then $|y(n)| > |y(O_k)|$. Points indexed by number series $J_k, k \geq 2$, also appear in the ascending branch, and they approach the horizontal axis with increasing $k$.

**Proof:** The approximation we obtained with the method of continued fraction $\tilde{\lambda}_k \approx \dfrac{O_k}{O_{k+1}}$ is the best approximation to the silver number $\lambda=\sqrt{2}-1$ [A4].

According to eq.(2) in the text, the point labeled with 0 is on the ascending branch, but point 1 is not. Point 2 is on the ascending branch since $|2\lambda - 1| \leq 1/4$. Based on the theory of best approximation, the absolute value $|O_k\lambda - O_{k-1}| < |2\lambda - 1| \leq 1/4$ for $k \geq 3$. So all points labeled with a number in the series $O_k$ for $k \geq 2$ appear in the ascending branch of the function $y = \arcsin(\sin(2\pi\lambda n))$.

Point 3 falls on the ascending branch since $|3\lambda - 1| \leq 1/4$. According to eq.(26), the absolute value $|J_{k+1}\lambda - J_k| \leq |J_2\lambda - J_1| = |3\lambda - 1| \leq 1/4$ for $k \geq 2$. So the points indexed with the number series $J_{k+1}$ for $k \geq 2$ appear in the ascending branch of the function,

and they approach the horizontal axis with increasing k.

**Theorem 3:** Points labeled with the numbers $O_0, O_k, O_{k+1}, J_{k+1}$ ($k \geq 2$) form a unit parallelogram in the plot of the ascending branch of the function $y = \arcsin(\sin(2\pi\lambda n))$.

**Proof:** From Theorem 2, we know that the points labeled with number $O_0, O_k, O_{k+1}, J_{k+1}$ ($k \geq 2$) are on the plot of the ascending branch. Let the ratio of the longitudinal scale to the transverse scale be $\gamma_k$. Then, the slope of the straight line joining the origin $O_0$ and point $O_k$ is

$$\beta_{O_0 O_k} = y(O_k)/x(O_k). \tag{28}$$

The slope of the straight line joining point $O_{k+1}$ and point $J_{k+1}$ is

$$\beta_{O_{k+1} J_{k+1}} = [y(J_{k+1}) - y(O_{k+1})]/[x(J_{k+1}) - x(O_{k+1})]. \tag{29}$$

From eqs.(24), (25) and (26), we can get that $x(J_{k+1}) = J_{k+1}\gamma = x(O_k) + x(O_{k+1})$ and $y(J_{k+1}) = 2\pi(J_{k+1}\lambda - J_l) = y(O_k) + y(O_{k+1})$. Putting these expressions into eq.(29), we get

$$\beta_{O_{k+1} J_{k+1}} = y(O_k)/x(O_k) = \beta_{O_0 O_k} \tag{30}$$

In a similar way we can prove that $\beta_{O_k J_{k+1}} = \beta_{O_0 O_{k+1}}$. Hence, a conclusion can be drawn that the points labeled with number $O_0, O_k, O_{k+1}, J_{k+1}$ ($k \geq 2$) form a parallelogram in the plot of the ascending branch of the function $y = \arcsin(\sin(2\pi\lambda n))$.

**Corollary:** There is no other point labeled with number less than $J_{k+1}$, but point $O_{k+1}$, whose corresponding value of the function $y = \arcsin(\sin(2\pi\lambda n))$ is less than that of $J_{k+1}$ ($k \geq 2$).

**Proof:** According to theorem 2, besides the origin there is no other point labeled with number less than $O_k$ whose corresponding value of the function $y = \arcsin(\sin(2\pi\lambda n))$ is less than that of point $O_k$. This is also true for point $O_{k+1}$. And according to theorem 3, those points labeled with number $O_0, O_k, O_{k+1}, J_{k+1}$ ($k \geq 2$) form a parallelogram. Such a parallelogram can be a unit, i.e., no other point falls in this parallelogram. From eq.(25) and eq.(26), we get that $|y(O_{k+1})| < |y(J_{k+1})| < |y(O_k)|$. So there is no other point labeled with number less than $J_{k+1}$, but the point $O_{k+1}$, whose corresponding value of the function $y = \arcsin(\sin(2\pi\lambda n))$ is less than that of $J_{k+1}$ ($k \geq 2$).

The lattice defined by the points in the ascending (descending) branch of the function $y = \arcsin(\sin(2\pi\lambda n))$ can be only approximately a square lattice. To verify this point, we compare the distances of points $O_k$, $O_{k+1}$ and $J_{k+1}$ ($k \geq 2$) to the origin. Let the ratio of the longitudinal scale to the transverse scale be $\gamma = \gamma_k$ when the distance of point $O_k$ to the origin is set equal to that of point $O_{k+1}$, i.e., $d_k(O_0 O_k) = d_k(O_0 O_{k+1})$, we get

$$d_k^2(O_0 O_k) = (O_k \gamma_k)^2 + [2\pi(O_k \lambda - O_{k-1})]^2 = O_k^2 \gamma_k^2 + 4\pi^2 \lambda^{2k} \tag{31}$$

$$d_k^2(O_0 O_{k+1}) = (O_{k+1} \gamma_k)^2 + [2\pi(O_{k+1} \lambda - D_k)]^2 = O_{k+1}^2 \gamma_k^2 + 4\pi^2 \lambda^{2k+2} \tag{32}$$

$$d_k^2(O_0 J_{k+1}) = (J_{k+1} \gamma_k)^2 + [2\pi(J_{k+1} \lambda - J_k)]^2 = J_{k+1}^2 \gamma_k^2 + 4\pi^2 \lambda^{2k}(1-\lambda)^2 \tag{33}$$

From eqs.(31) and (32), and the equality $d_k(O_0 O_k) = d_k(O_0 O_{k+1})$, and the fact $\lambda = \sqrt{2} - 1$, we get

$$\gamma_k^2 = \frac{8\pi^2 \lambda^{2k+1}}{\left(O_{k+1}^2 - O_k^2\right)} \tag{34}$$

Putting the expression of $\gamma_k^2$ back into eq.(31) and eq.(32), we get

$$d_k^2(O_0 O_k) = d_k^2(O_0 O_{k+1}) = 4\pi^2 \lambda^{2k} \frac{\left[O_{k+1}^2 - \lambda O_k^2\right]}{\left(O_{k+1}^2 - O_k^2\right)} \tag{35}$$

$$d_k^2(O_0 J_{k+1}) = 8\pi^2 \lambda^{2k} \frac{\left[\left(O_{k+1}^2 + O_k O_{k+1}\right) - \lambda^2\left(O_k^2 + O_k O_{k+1}\right) - \lambda\left(O_{k+1}^2 - O_k^2\right)\right]}{\left(O_{k+1}^2 - O_k^2\right)} \tag{36}$$

For a rigorous square with a side length $d_k(O_0O_k)$, the squared length of the diagonal should be

$$\left[\sqrt{2}d_k(O_0O_k)\right]^2 = 2d_k^2(O_0O_k) = 8\pi^2\lambda^{2k}\frac{\left(O_{k+1}^2 - \lambda^2 O_k^2\right)}{\left(O_{k+1}^2 - O_k^2\right)} \tag{37}$$

The difference between $\left[\sqrt{2}d_k(O_0O_k)\right]^2$ and $d_k^2(O_0J_{k+1})$ is thus

$$\begin{aligned}\delta_k &= d_k^2(O_0J_{k+1}) - 2d_k^2(O_0O_k) = \frac{8\pi^2\lambda^{2k+1}}{\left(O_{k+1}^2 - O_k^2\right)}\left[-O_{k+1}^2 + 2O_kO_{k+1} + O_k^2\right] \\ &= (-1)^{k-1}\frac{8\pi^2\lambda^{2k+1}}{\left(O_{k+1}^2 - O_k^2\right)} = (-1)^{k-1}\gamma_k^2\end{aligned} \tag{38}$$

In the deduction process above we have employed the equality $O_{k+1}^2 - 2O_kO_{k+1} - O_k^2 = (-1)^k$.

Thus, it can be concluded that points generated by the ascending (descending) branch of the function $y = \arcsin(\sin(2\pi\lambda n))$ form an approximate square lattice when a proper value for $\gamma$ is chosen.

The relative deviation of the diagonal length can be defined as

$$\eta_l = \frac{|\delta_k|}{d_k^2(O_0O_k)} = \frac{2\lambda}{\left(O_{k+1}^2 - \lambda^2 O_k^2\right)} \tag{39}$$

Since $\delta_k$ may change sign with varying $k$, here the absolute value $|\delta_k|$ is adopted.

This deviation satisfies the following inequality

$$\eta_k = \frac{2\lambda}{\left(O_{k+1}^2 - \lambda^2 O_k^2\right)} < \frac{2\lambda}{\left(O_{k+1}^2 - O_k^2\right)} = \frac{2\lambda}{\left(3O_k^2 + O_{k-1}^2 + 4O_{k-1}O_k\right)} \tag{40}$$

Since the value of $O_k$ increases with increasing k, so does $\eta_l$. At $k \to \infty$, $\eta_l(k)=0$, thus the plot of the ascending branch of the function $y = \arcsin(\sin(2\pi\lambda n))$ forms a rigorous square lattice. The same conclusion can be drawn for the descending branch.

The scaling at $k \to \infty$ is then operated on a rigorous square lattice, which reveals the presence of the directional scaling symmetry (along the horizontal direction in

Fig.S2) for the square lattice.

The connection line between point 0 and point 17 in Fig.5a, and the connection line between point 0 and point 29 in Fig. 5b, are roughly at 22.5°, or $\pi/8$, with respect to the horizontal axis. Calculation below shows that the direction of scaling symmetry for the square lattice, going through a lattice point, lies at 22.5° with regard to the side of the unit square.

Combining $y(O_{k+1}) = 2\pi(O_{k+1}\lambda - O_k) = (-1)^k 2\pi\lambda^{k+1}$, the eq.(35) and the fact

$\sin(22.5°) = \dfrac{\sqrt{2-\sqrt{2}}}{2}$ (or $\sin^2(22.5°) = \dfrac{2-\sqrt{2}}{4}$), one obtains

$$y^2(O_{k+1}) - d_k^2(O_0 O_{k+1})\sin^2(22.5°) = (2\pi\lambda^{k+1})^2 - 4\pi^2\lambda^{2k}\dfrac{(O_{k+1}^2 - \lambda^2 O_k^2)}{(O_{k+1}^2 - O_k^2)}\dfrac{2-\sqrt{2}}{4}$$
$$= 4\sqrt{2}\pi^2\lambda^{2k+1}\left[\dfrac{\lambda^2 O_{k+1}^2 - O_k^2}{4(O_{k+1}^2 - O_k^2)}\right] \quad (41)$$

Defining a relative deviation from the square as

$$\Delta_k = \dfrac{y^2(O_{k+1}) - d_k^2(O_0 O_{k+1})\sin^2(22.5°)}{d_k^2(O_0 O_{k+1})} = \dfrac{\sqrt{2}\lambda(\lambda^2 O_{k+1}^2 - O_k^2)}{4(O_{k+1}^2 - \lambda^2 O_k^2)} \quad (42)$$

As at $k \to \infty$, $O_{k+1}/O_k \to \sqrt{2}+1$, hence $\Delta_k \to 0$. Thus the direction of scaling symmetry, going through a lattice point, for the square lattice is at 22.5° with respect to the side of the unit square.

From eq.(34), we can easily obtain the squared value for $\gamma = \gamma_{k+1}$, or the ratio of the longitudinal scale to the transverse scale when $d_{k+1}(O_0 O_{k+1}) = d_{k+1}(O_0 O_{k+2})$,

$$\gamma_{k+1}^2 = \dfrac{8\pi^2\lambda^{2k+3}}{(3O_{k+1}^2 + 4O_k O_{k+1} + O_k^2)} \quad (43)$$

Combining eqs. (25) and (43), one obtains the ratio of $\gamma_{k+1}^2$ over $\gamma_k^2$,

$$\frac{\gamma_{k+1}^2}{\gamma_k^2} = \frac{\lambda^2(O_{k+1}-O_k)}{(3O_{k+1}+O_k)} \tag{44}$$

Thus it has

$$\lim_{k\to\infty}\frac{\gamma_{k+1}}{\gamma_k} = \lambda^2 = 3-2\sqrt{2} \tag{45}$$

Which is the scaling factor involved in the directional scaling symmetry for the square lattice.

From eq.(35), it is easy to get the squared distance from $O_{k+1}$ to the origin at $\gamma = \gamma_{k+1}$

$$d_{k+1}^2(O_0O_{k+1}) = 4\pi^2\lambda^{2k+2}\frac{\left[(1+2\sqrt{2})O_{k+1}^2+4O_kO_{k+1}+O_k^2\right]}{(3O_{k+1}^2+4O_kO_{k+1}+O_k^2)} \tag{46}$$

Combining eqs.(35) and (46), we can get the ratio of squared side lengths of the approximate unit squares under the condition $\gamma = \gamma_k$ and $\gamma = \gamma_{k+1}$, respectively,

$$\frac{d_{k+1}^2(O_0O_{k+1})}{d_k^2(O_0O_k)} = \lambda^2\frac{(1+2\sqrt{2})\left(O_{k+1}+\frac{3+\sqrt{2}}{7}O_k\right)(O_{k+1}-O_k)}{3\left[O_{k+1}+(1-\sqrt{2})O_k\right]\left(O_{k+1}+\frac{1}{3}O_k\right)} \tag{47}$$

Thus it has

$$\lim_{k\to\infty}\frac{d_{k+1}(O_0O_{k+1})}{d_k(O_0O_k)} = \lambda = \sqrt{2}-1 \tag{48}$$

For the square lattice, again the scale factor for directional scaling symmetry is the squared ratio of side lengths of unit squares before and after the contraction along the directional scaling axis.

In summary, for the square lattice, scaling symmetry appears at the direction, going through a lattice point, at 22.5° with respect to the side of the unit square, the scale factor is $3-2\sqrt{2}$, and the side length of the unit square is scaled by $\lambda = \sqrt{2}-1$. Such a contraction can be performed repeatedly.

**IV. Obtaining honeycomb lattice from the equilateral triangular lattice**

In order to obtain a honeycomb lattice by removing points from an equilateral triangular lattice as indexed in Fig.S1, the points to wipe off are chosen by the following rules:

(1) Choose a point with index $n=7a+12b$, where n, a, b are all integers, thus this point can be specified with (a, b). Remove this point;

(2) Next, remove the six points around point (a, b) given by (a+1, b+1), (a-1, b-1); (a+1,b), (a-1,b); and (a-2, b-1), (a+2,b+1). If the point has already been removed, ignore it;

(3) Repeat step (2) until half of all points are removed and a honeycomb lattice is obtained.